# $Sn_2Pd$: a possible superconducting material with topological surface states

M. M. Sharma[1,2], V.P.S. Awana[1,2]

[1]*National Physical Laboratory (CSIR), Dr. K. S. Krishnan Road, New Delhi 110012, India.*

[2]*Academy of Scientific and Innovative Research, Ghaziabad, U.P. 201002, India.*

**Abstract:**

In this article, we report the detailed magneto transport measurements of topological semimetal (TSM) candidate, $Sn_2Pd$. Single crystal of $Sn_2Pd$ is synthesized through self-flux method. Phase purity and crystalline morphology are confirmed through powder X ray diffraction (PXRD) pattern and field emission scanning electron microscopy (FESEM) image. Signatures of superconducting transition are seen in both transport and magneto transport measurements, which are further supported by the AC magnetization studies. $Sn_2Pd$ is found to show superconductivity onset at below 2.8K, but not the zero resistivity down to 2K. Although, isothermal magneto resistivity measurements below superconducting onset (2.8K) clearly exhibited superconductor like behavior, but is not observed in heat capacity measurements, indicating the same to be of weak nature. Magneto transport measurements performed in normal state of $Sn_2Pd$ show the presence of weak antilocalization (WAL) effect, which is confirmed by modelling of low field magneto-conductivity (MC) through Hikami Larkin Nagaoka (HLN) formalism. Here, it is worth mentioning that the present article is the first report on magneto transport measurements of $Sn_2Pd$, which show the same to be a topological material with weak superconducting phase below around 2.8K.

*Corresponding Author
Dr. V. P. S. Awana: E-mail: awana@nplindia.org
Ph. +91-11-45609357, Fax-+91-11-45609310
Homepage: awanavps.webs.com

## Introduction:

In recent years, the materials with exotic topological phases such as topological insulators (TIs), topological semimetals (TSMs) and topological superconductors (TSCs) have generated a significant interest of condensed matter scientists [1-3]. These topological materials contain exotic topological surface states (TSS), which allows electrical current to flow without any dissipation [1,2]. TSCs contain a full superconducting gap in bulk states along with topological surface states at the boundary, which are supposed to hosts Majorana fermions [3]. Till now, a lot of efforts have been made to fabricate TSCs, which include heterostructures of conventional superconductor and TI material or high spin orbit coupling (SOC) semiconducting materials [4-6]. Apart from the heterostructures, some of the doped TIs with formula $A_xBi_2(Te/Se)_3$ (A=Cu, Nb, Sr, Tl, Pd) [7-11] are found to show topological superconductivity. In regards of topological superconductivity, TSMs are extensively studied due to their high carrier density and intrinsic topological properties [12]. TSMs are interesting materials as the bulk states of these materials show topological non-trivial structure, which is only possible for surface states in case of TIs [2,12,13]. In TSMs, the valence and conduction bands are supposed to show band degeneracy at high symmetry points or lines in Brillouin zone [14] and are linearly dispersed [15]. Linear dispersion of bands in TSMs leads to occurrence of linear magnetoresistance (LMR) in magneto transport measurements [16-18]. TSMs possess high carrier density, which makes them as a prominent candidate to observe superconductivity with topological non-trivial band structure [12].

TSMs having the general formula $X_2Pd$ (X=Bi, Pb, Sb) are reported to have topological non-trivial band structure along with bulk superconductivity [19-29]. β-$Bi_2Pd$ is extensively studied for its topological and superconducting properties [19-21]. Monolayer of β-$Bi_2Pd$ shows evidence of existence of p wave superconductivity, and also the possible signatures of Majorana zero modes [30,31]. The other members of this family viz. $Pb_2Pd$ and $Sb_2Pd$ are lesser studied [22-29]. $Sb_2Pd$ is superconducting below 1.35K with topological non-trivial band structure [22-26]. Another compound $Pb_2Pd$ is also superconducting and the theoretical calculations show the same to be a topological material [27-29]. Simultaneous existence of superconductivity with topological non-trivial band structure makes this family of compounds interesting. $Sn_2Pd$ also belongs to this family of materials, and remained unexplored in context of topological properties. This is despite an early report based on theoretical studies of band structure of $Sn_2Pd$ showing the same to be a TSM [14]. Magneto-transport properties of $Sn_2Pd$ remained unexplored till date as the one and only report available is on the R-T measurements of the same [32]. Lack of experimental studies on $Sn_2Pd$, motivates us to study its magneto transport and magnetic properties in order to probe possible superconductivity in this TSM.

This article reports the aspects of superconductivity and topological properties of $Sn_2Pd$ single crystal. Single crystal of $Sn_2Pd$ is synthesized through a self-flux method using solid state reaction route. Glimpses of superconductivity are observed in both transport and magnetic measurements. Interestingly, synthesized $Sn_2Pd$ single crystal is observed to show non saturating LMR, signifying the possible existence of TSS in the studied system. Low field MC data is further analysed using HLN model, which confirmed the presence of WAL effect and shows that the conduction is dominated by TSS at low temperatures.

**Experimental:**

$Sn_2Pd$ single crystal is grown by using a simple self-flux method. Sn and Pd were taken in stoichiometric ratio (2:1) and grounded in Argon atmosphere inside MBraun made Glove Box. The obtained homogenous powder was then palletized and vacuum encapsulated in quartz ampoule. The sample is then heated to 950°C for 24 hours and then slowly cooled to 580°C at a rate of 2°C/h. The sample was kept at this temperature for 48 hours and then quenched in ice water. The obtained crystal was silvery shiny. The schematic of heat treatment is shown in Fig. 1(a). Rigaku mini flex II table top XRD equipped with Cu-$K_\alpha$ radiation is used to record PXRD pattern. Surface morphology and elemental composition were analysed using MIRA II LMH TESCAN made FESEM equipped Energy Dispersive X-Ray (EDX) detector. Transport, magnetic and thermal measurements were performed using Quantum Design Physical Property Measurement System (QD-PPMS) equipped with ±14T magnet. Full Prof software was used for Rietveld refinement of PXRD data and VESTA software was used to draw the unit cell of synthesized $Sn_2Pd$ single crystal.

**Results & Discussion:**

Fig. 1(b) depicts Rietveld refined PXRD pattern of synthesized $Sn_2Pd$ single crystal. Rietveld refinement is performed using the parameters of a tetragonal unit cell with I 41/a c d space group symmetry. The PXRD data is found to be well fitted with the applied fitting protocols and the parameter of goodness of fit i.e., $\chi^2$ is found to be 2.20, which is in an acceptable range. All peaks of PXRD pattern are found to be fitted with applied space group symmetry and the parameters of tetragonal phase, and are indexed with their respective planes. The fitted PXRD pattern shows that the sample is phase pure and no secondary phase is present in the system. Also, peaks for individual elemental impurity are not detected in PXRD pattern, which shows that no element remained unreacted. The obtained lattice parameters from Rietveld refinement are a=b=6.492(4)Å & c=24.363(2)Å, $\alpha=\beta=\gamma=90°$. The unit cell parameters along with the fitting parameters are given in table-1. The crystallographic information file (CIF) generated from Rietveld refinement is used to draw unit cell of synthesized $Sn_2Pd$ single crystal using VESTA software, and the same is shown in inset of Fig. 1(b). A single unit cell of $Sn_2Pd$ is found to have 32 atoms of Sn and 16 atoms of Pd making the atomic formula to be $Sn_{32}Pd_{16}$. Single crystalline nature of synthesized $Sn_2Pd$ single crystal is confirmed through FESEM image, which is shown in inset of Fig. 2(a). A typical slab like morphology can be seen in FESEM image, which shows that the synthesized sample is crystalline in nature. This slab like morphology signifies the layer-by-layer growth of the synthesized $Sn_2Pd$ single crystal. Purity of the sample is verified through EDX measurement, which is shown in fig. 2(a). EDX spectra contains the peaks only for the constituent elements viz. Sn and Pd, which shows that no impurity of foreign atoms is present in the system. The composition of synthesized $Sn_2Pd$ single crystal, obtained from EDX measurement is found to be $Sn_{2.1}Pd$. EDX mapping is also performed to check the distribution of constituent elements viz. Sn and Pd throughout the sample and is shown in fig. 2(b) and 2(c) respectively. Both the elements are found to be homogenously distributed throughout the sample. Homogenous distribution of Sn and Pd shows that none of the element remained in unreacted form in the synthesized sample, which is in agreement with the PXRD results. Both PXRD and EDX measurements shows that the synthesized $Sn_2Pd$ single crystal is phase pure and no unreacted impurity is present in the sample.

$Sn_2Pd$ has the general formula similar to materials $Bi_2Pd$, $Sb_2Pd$ and $Pb_2Pd$, which all are superconducting in nature. Superconducting properties of $Sn_2Pd$ have not been revealed yet, so transport measurements have been carried out on synthesized $Sn_2Pd$ single crystal in a temperature range from 100K to 2K as shown in fig. 3(a). $Sn_2Pd$ shows metallic behaviour as resistivity decreases with decrease in the temperature, and the same becomes almost constant at below 10K showing Fermi liquid behaviour. Resistivity data is fitted with the following formula

$$\rho(T) = \rho_0 + AT^2 \tag{1}$$

Here $\rho_0$ represents residual resistivity arising due to impurity scattering and is found to be 5.4mΩ-cm. The residual resistivity ratio (RRR) i.e., $\rho_{100}/\rho_5$ is found to be 1.06, which is rather low suggesting either the presence of structural disorder or low carrier concentration in the synthesized $Sn_2Pd$ single crystal. The fitted plot is shown in the left inset of the fig. 3(a). Resistivity is well fitted with the applied formula up to 30K and starts to deviate above it, showing electron-electron (e-e) scattering to be the dominant scattering till 30K and electron-phonon (e-p) scattering dominating at higher temperatures. Interestingly, resistivity drops sharply below 2.8K, which can be attributed as a possible superconducting transition. At this particular point resistivity is decreased almost 74%, but this transition does not reach to zero resistivity down to 2K (lowest temperature limit of instrument). A similar superconducting transition has also been observed in $TaIrTe_4$ [33], which is also a TSM. In $TaIrTe_4$ also, zero resistivity was not achieved, while resistivity was dropped almost 44% from the residual resistivity value. Here also, zero resistivity is not achieved but the resistivity is decreased significantly, hinting towards possible superconducting transition in synthesized $Sn_2Pd$ single crystal. For the sake of repeatability, transport measurements were carried out on five different samples and all of them showed significant drop in resistivity, thus exhibiting the consistency of the occurrence of superconductivity. Sample S3, showed the maximum drop of around 74% as shown in fig. 3(a), and this sample is considered for other measurements. Further, the magneto transport (ρ-H) measurements have been carried out to check that the observe resistivity drop is a superconducting transition or not, as for a superconducting transition, upper critical field should be lowered as the temperature is increased. Fig. 3(b) shows ρ-H plot of synthesized $Sn_2Pd$ single crystal, at 2K, 2.1K, 2.2K, 2.3K, 2.4K, 2.5K, 2.6K, 2.7K and 3K. Here also, zero resistivity is not observed below the critical field as observed in ρ-T plot in fig. 3(a). It is clear form fig. 3(b), that the sample is in superconducting state below 2.7K, as the resistivity is constant up to a certain magnetic field and the same is found to increase sharply when the field is increased further. This particular point, where resistivity jumps sharply is termed as critical field ($H_c$). Resistivity drop is not observed in ρ-H plot at 3K, showing the sample is clearly in normal state. $H_c$ is shifted toward lower field value as the temperature is increased as can be seen in fig. 3(b), this is a typical behaviour of a conventional superconductor. The value of $H_c$ at 2K is found to be around 270Oe. The observed transition in both ρ-T and ρ-H measurements cannot be attributed to the transition due to superconducting impurity of Sn, as the same is superconducting at 3.6K, which is much higher than the transition temperature observed in the present case. The presence of structural disorder can be the possible reason that zero resistivity is not achieved up to 2K and a larger transition width is observed. This is consistent with the observed low value of RRR. Presence of structural disorders in a superconducting sample leads to breaking of superconducting channels, due to which resistivity does not drop to zero value. In literature, non-occurrence of zero resistivity in a superconducting transition is attributed to possible quasi-1D superconductivity or surface superconductivity [33,34], in which only the helical surface states become superconducting. In quasi-1D superconductors, non-vanishing resistivity is observed in superconducting state due to phase slip events [35].

Fig. 4 shows AC magnetization measurements of synthesized $Sn_2Pd$ single crystal at various AC fields viz. 3Oe, 5Oe, 7Oe, 9Oe, 11Oe and 13Oe. AC field frequency is set to 333Hz and the background DC field was set to 0Oe, throughout the measurement. The upper plot in fig. 3(b), is showing the imaginary part of AC magnetic moment (M'') while the lower plot is showing the real part (M') of the same. A superconducting transition is visible with $T_c^{onset}$ at 2.8K, in both parts of AC magnetic moment. None of the moment saturates down to 2K, confirming superconducting volume fraction to be low, which is consistent with occurrence of non-zero resistivity in transport measurements. In both of the measurements, viz. transport and AC magnetization, the superconducting transition does not saturate, which shows that the observed superconducting transition is not the bulk property of the sample. At the same time, the observed diamagnetic signal in AC magnetization measurements opposes the possibilities of the superconducting transition to be termed as surface superconductivity. Heat capacity measurements are also carried out to check whether the observed superconductivity is bulk property of the sample or not. Fig. 5 shows the heat capacity measurements of synthesized $Sn_2Pd$ single crystal, from 195K to 2K. Superconducting transition is not observed down to 2K, signifying the absence of bulk superconductivity in $Sn_2Pd$. Interestingly, although the observed Meissner effect in AC magnetization measurements suggest that the observed superconductivity is not due to the surface states, yet the observed low diamagnetic signal and non-occurrence of superconductivity in heat capacity measurements suggest that it is not the bulk superconductivity either. These results are similar to other superconducting topological material e.g. Pd doped $Bi_2Te_3$ [36-39]. Pd doped $Bi_2Te_3$ also shows low superconducting volume fraction and non-occurrence of zero resistivity, yet its superconductivity is not attributed to any superconducting impurities present in the sample [36]. Similarly, some other topological materials viz. Cu doped $Bi_2Se_3$ [36,40] and Ti doped $NiTe_2$ [41] shows finite resistivity and a low diamagnetic signal in the superconducting state. Here, $Sn_2Pd$ also follows the same trend as the above-mentioned topological materials do, and the presence of structural disorders can be the possible reason for the observed properties. This warrants more experimental work to be performed on this system to further explore its structural and superconducting properties.

Further, specific heat data ($C_p$ vs T) is utilized to determine the normal state parameters of synthesized $Sn_2Pd$ single crystal. For this, low temperature $C_p/T$ is plotted against $T^2$ as shown in inset of fig. 5. Heat capacity of a material is contributed by two terms viz. electronic contribution to specific heat ($C_{el}$) and phonon contribution to specific heat ($C_{ph}$). Total specific heat is given by the following formula:

$$\frac{C_p}{T} = \gamma_n + \beta_n T^2 \tag{2}$$

here $\gamma_n T$ represents $C_{el}$ and $\beta_n T^3$ represents $C_{ph}$. The coefficient associated with these terms are determined by fitting the $C_p/T$ vs $T^2$ plot with linear equation as shown in inset of fig. 5. The intercept of linear fit provides the value of $\gamma_n$, which is known as Sommerfeld coefficient and the slope of the same provides the value of $\beta_n$, which is used in determining the value of Debye temperature ($\theta_D$). The obtained values of $\gamma_n$ and $\beta_n$ are found to be 5.14 ± 0.35 mJ mol$^{-1}$ K$^{-2}$ and 1.13 ± 0.02 mJ mol$^{-1}$ K$^{-4}$. Sommerfeld coefficient is used to determine the Density of States at the Fermi level i.e., $D_c(E_F)$, which is given by the following formula,

$$\gamma_n = \frac{\pi^2 k_B^2 D_c(E_F)}{3} \tag{3}$$

The obtained value of $D_c(E_F)$ is found to be 2.19 states eV$^{-1}$f.u.$^{-1}$. $\theta_D$ is related to $\beta_n$ with the following formula

$$\theta_D = \left(\frac{12\pi^4 nR}{5\beta_n}\right)^{1/3} \tag{4}$$

Here R=8.314 J mol$^{-1}$ K$^{-2}$ and n is the number of atoms in the formula unit, which is 3 for $Sn_2Pd$. The obtained value of $\theta_D$ for synthesized $Sn_2Pd$ single crystal is 172.2K.

Normal state properties of $Sn_2Pd$ are determined through magneto-transport measurements at temperatures above the critical temperature, and the same is shown in fig. 6(a). The variation in resistivity in a field range of ±12T, at various temperatures viz. 5K, 10K, 20K, 30K, 40K, 50K has been calculated in terms of MR% using the following formula:

$$MR\% = \left[\frac{\rho(H) - \rho(0)}{\rho(0)}\right] \times 100 \tag{5}$$

A significant MR% of around 50% has been observed at 5K, which gradually decreases with increasing the temperature. In normal metals, MR can arise due to change in trajectory of electrons on application of magnetic field, which has quadratic dependency on magnetic field. This behaviour is more pronounced at lower magnetic field. To check, whether the observed MR is normal one, which occurs in non-magnetic metallic systems, the low field MR% (upto 4T) has been plotted against $H^2$, as shown in inset of fig. 6(a). It is clear from inset of fig. 6(a) that low field MR is not linear with $H^2$, suggesting that the observed MR in $Sn_2Pd$ is not normal metallic MR. The reason for occurrence of MR in topological materials is different and is related to $\pi$ Berry phase of TSS. The presence of $\pi$ Berry phase results in destructive interference between two time-reversed paths of electrons. This destructive interference results in delocalization of electrons, which is known as WAL effect. Application of magnetic field disturbs the $\pi$ Berry phase, and results in occurrence of positive MR in topological materials. The WAL induced MR shows cusp like behaviour at low magnetic field [42,43] as observed in present study on $Sn_2Pd$ single crystal. In topological materials WAL arises due to the presence of topological surface state, while some materials with high spin orbit coupling also shows WAL effect [42-47].

The fitted ρ-T plot shows the presence of both e-e and e-p scattering processes, this information can be verified by applying Kohler scaling laws in MR data. Kohler's rule suggests that MR% should be a function of Hτ, where τ is the time spent in scattering events of conduction electrons. τ is inversely proportional to zero field resistivity ρ(0), so MR% should be a function of $H/\rho_0$. If a system preserves the Kohler's law, there exists only a single e-e scattering and MR% vs $H/\rho_0$ plots at different temperatures should collapse into each other [48-50]. To check this, MR% is plotted against $H/\rho_0$ and plotted in fig. 6(c). It is clear from fig. 6(c), that Kohler rule is violated in $Sn_2Pd$ single crystal as all the MR% vs $H/\rho_0$ plots at different temperatures do not merge into a single curve. This violation of Kohler's law suggests that two scattering processes viz. e-e and e-p scattering are present in the synthesized $Sn_2Pd$ single crystal. This is further verified by calculation of phase coherence length $l_\phi$ at different temperatures.

Fig. 6(c) shows the low field (±0.6T) MC data of synthesized $Sn_2Pd$ single crystal at various temperatures viz. 5K, 10K, 20K, 30K, 40K & 50K. The low field MC data is fitted with HLN equation [51], which is as follows

$$\Delta\sigma(H) = -\frac{\alpha e^2}{\pi h}\left[ln\left(\frac{B_\varphi}{H}\right) - \Psi\left(\frac{1}{2} + \frac{B_\varphi}{H}\right)\right] \quad (6)$$

Here $\Delta\sigma(H)$ symbolizes $[\sigma(H)-\sigma(0)]$, $B_\phi$ is termed as characteristic field, which is given by $B_\varphi = \frac{h}{8e\pi l_\varphi^2}$, Ψ is digamma function. $l_\phi$ denotes the phase coherence length, which is the maximum length, which is travelled by the electron while maintaining its phase. Another important parameter, which is obtained from HLN fitting is the pre-factor α. The value of α confirms that whether the WAL effect is present in the system or not, as a negative value of α signifies the presence of WAL effect, while a positive value of the same shows the presence of weak localization (WL) effect. The value of $\alpha$, gives the information about the conduction mechanism, whether the conduction is solely dominated by surface states or the bulk states also contribute in conduction mechanism. $\alpha$ is supposed to take value -0.5 for single TSS [42], while for two independent TSS the value of $\alpha$ should be equal to -1. The deviation of $\alpha$ from -0.5 and -1, signifies that the TSS are connected through bulk states [52-54], and the bulk states also contribute to conduction mechanism. HLN fitted low field (±0.6T) MC data at 5K, 10K, 20K, 30K, 40K and 50K is shown by solid black line in Fig. 6(c). The value of pre-factor $\alpha$ at 5K is found to be -0.56, the negative value of $\alpha$ confirms the presence of WAL effect in synthesized $Sn_2Pd$ single crystal. The obtained value of $\alpha$ is close to -0.5 suggesting that single conducting channel is present in the system, while its deviation from standard value of -0.5, suggests that the bulk states also contribute in the conduction mechanism. The value of $\alpha$ tends to decrease monotonically as the temperature is increased, as shown in fig. 6(d). The value of $\alpha$ is very close to standard value -0.5 at 5K and 10K, suggesting that surface states dominate in conduction mechanism at both of these temperatures. Further, at higher temperatures viz. 20K, 30K, 40K and 50K, the value of $\alpha$ is deviated largely from -0.5 significantly, suggesting that the conduction phenomenon is mainly governed by the bulk states at these temperatures. Phase coherence length $l_\phi$, is also calculated through HLN modelling, and the variation in $l_\phi^{-2}$ with respect to temperature is shown in fig. 6(d) by blue symbols. The variation of $l_\phi$ with respect to temperature gives important information about the dephasing mechanism and scattering processes. The Nyquist criterion suggests that only a single scattering process i.e. electron-electron (e-e) scattering is included in dephasing mechanism if $l_\phi^{-2}$ varies linearly with temperature [55,56]. A deviation of $l_\phi^{-2}$ vs T plot from linearity suggests that there exists another scattering process, which is electron-phonon (e-p) scattering. It is clear from fig. 6(d), that the variation in $l_\phi^{-2}$ is not linear with respect to temperature, suggesting the contribution of both e-e and e-p scattering process in dephasing mechanism in $Sn_2Pd$ single crystal. The involvement of both the scattering processes is in accordance with the violation of Kohler rule in MR vs $H/\rho_0$ plot. Further, the $l_\phi^{-2}$ vs T plot is fitted with the following power law [57] shown by solid black line in fig. 6(d):

$$\frac{1}{l_\Phi^2(T)} = \frac{1}{l_\Phi^2(0)} + A_{e-e}T^p + A_{e-p}T^q \tag{7}$$

Here $l_\phi(0)$ represents the dephasing length at absolute zero and the contribution from e-e and e-p scattering are shown by $A_{e-e}T^p$ and $A_{e-p}T^q$ respectively. The value of p and q from fitting are found to be 1 and 2 respectively, which shows the 2-D nature of observed WAL effect, which is observed in other topological materials [54,58]. The obtained value of $l_\phi(0)$ from power law fitting is 102nm, and the values of $A_{e-e}$ and $A_{e-p}$ are found to be $2.185\times10^{-6}$ and $3.32\times10^{-8}$. The values of pre factor $\alpha$ and dephasing length and phase coherence length $l_\phi$ at various temperatures obtained from HLN fitting are listed in table-2.

**Conclusion:**

Summarily, this article reports a detailed report on synthesis and magneto transport measurements of an unexplored topological material, $Sn_2Pd$. Phase purity of stoichiometry is evidenced through XRD and EDX measurements. $Sn_2Pd$ is found to show evidences of possible superconducting transition in the form of significant resistivity drop in both ρ-T and ρ-H measurements, while zero resistivity is not achieved. Superconducting transition is not observed in heat capacity measurements, which further shows the absence of bulk superconducting properties. A non-saturating LMR is observed in magneto transport measurements, showing the possible presence of TSS, which is further confirmed by HLN fitting of low field MC data. This is ***the first report on magneto transport measurements*** showing the experimental evidence of topological properties along with the low temperature (below 2.8K) surface superconductivity in $Sn_2Pd$ single crystal.

**Acknowledgment:**

Authors would like to thank Director of National Physical Laboratory, New Delhi for his keen interest. M. M. Sharma would like to thank CSIR, India for research fellowship and AcSIR, Ghaziabad for Ph.D. registration.

**Table-1**

Parameters obtained from Rietveld refinement:

| Cell Parameters | Refinement Parameters |
|---|---|
| Cell type: Tetragonal<br>Space Group: I 41/acd<br>Lattice parameters: a=b=6.492(4)Å<br>& c=24.362(2)Å<br>α=β=γ=90° | $\chi^2$=2.20<br>$R_p$=7.37<br>$R_{wp}$=9.66<br>$R_{exp}$=6.51 |
| Cell volume: 1026.959Å$^3$<br>Density: 8.894 g/cm$^3$<br>Atomic co-ordinates:<br>Sn1 (0.15785,0.40785,0.125)<br>Pd (0,0.25,0.31665)<br>Sn2 (0.24741,0,0.25) | |

**Table: 2**

Low field (up to 1 Tesla) HLN fitted parameters of $Sn_4Au$ single crystal

| Temperature(K) | $\alpha$ | $l_\phi$(nm) |
|---|---|---|
| 5 | -0.5608 | 95.06 |
| 10 | -0.4749 | 91.8619 |
| 20 | -0.2694 | 83.7321 |
| 30 | -0.2335 | 72.0596 |
| 40 | -0.2319 | 63.9714 |
| 50 | -0.1986 | 59.5581 |

**Figures Caption:**

**Fig. 1(a):** Schematic of heat treatment used to synthesize $Sn_2Pd$ single crystal.

**Fig. 1(b):** Rietveld refined PXRD pattern of synthesized $Sn_2Pd$ single crystal in which the red symbols are showing the experimental data and the solid black line is showing the Rietveld refined data, the inset is showing the unit cell of the same drawn by using VESTA software where blue circles are showing the Pd atoms and the orange circles are showing the Sn atoms.

**Fig. 2(a):** EDX spectra showing the peaks of constituent elements of $Sn_2Pd$ single crystal in which inset is showing the FESEM image of the same. **(b)** EDX mapping of Sn atoms **(c)** EDX mapping of Pd atoms.

**Fig. 3(a):** $\rho$ vs T plot of synthesized $Sn_2Pd$ single crystal in temperature range from 100K to 2K, in which right inset is showing the zoomed view of $\rho$ vs T plot in proximity of superconducting transition and the left inset is showing the fitted $\rho$ vs T plot in normal state of $Sn_2Pd$ from 5K to 100K.

**Fig. 3(b):** $\rho$ vs H plot of synthesized $Sn_2Pd$ single crystal at temperatures 2K, 2.1K, 2.2K, 2.3K, 2.4K, 2.5K, 2.6K, 2.7K and 3K.

**Fig. 4:** AC magnetization measurements of $Sn_2Pd$ single crystal at various AC fields viz. 3Oe, 5Oe, 7Oe, 9Oe, 11Oe & 13Oe.

**Fig. 5:** $C_p$ vs T plot of synthesized $Sn_2Pd$ single crystal, in which inset is showing the fitted $C_p/T$ vs $T^2$ plot of the same.

**Fig. 6(a):** MR% vs H plot in a field range of ±12T at various temperatures viz. 5K, 10K, 20K, 30K, 40K and 50K, in which inset is showing the variation in MR% with respect to square of applied field upto 4T.

**Fig. 6(b):** MR% vs $H/\rho_0$ plots for Kohler's scaling of MR at various temperatures viz. 5K, 10K, 20K, 30K, 40K and 50K.

**Fig. 6(c):** HLN fitted low field (±12T) MC data of $Sn_2Pd$ single crystal at 5K, 10K, 20K, 30K, 40K and 50K, in which the HLN fitted plots are shown by solid black lines.

**Fig. 6(d):** Variation in parameters obtained from HLN fitting with respect to temperature in which red symbols are showing the variation in pre factor $\alpha$ with respect to temperature and the blue symbols are showing the same for inverse of square of phase coherence length ($l_\phi^{-2}$). The solid black line is showing the power law fitted $l_\phi^{-2}$ vs T plot of synthesized $Sn_2Od$ single crystal

Fig. 1(a)

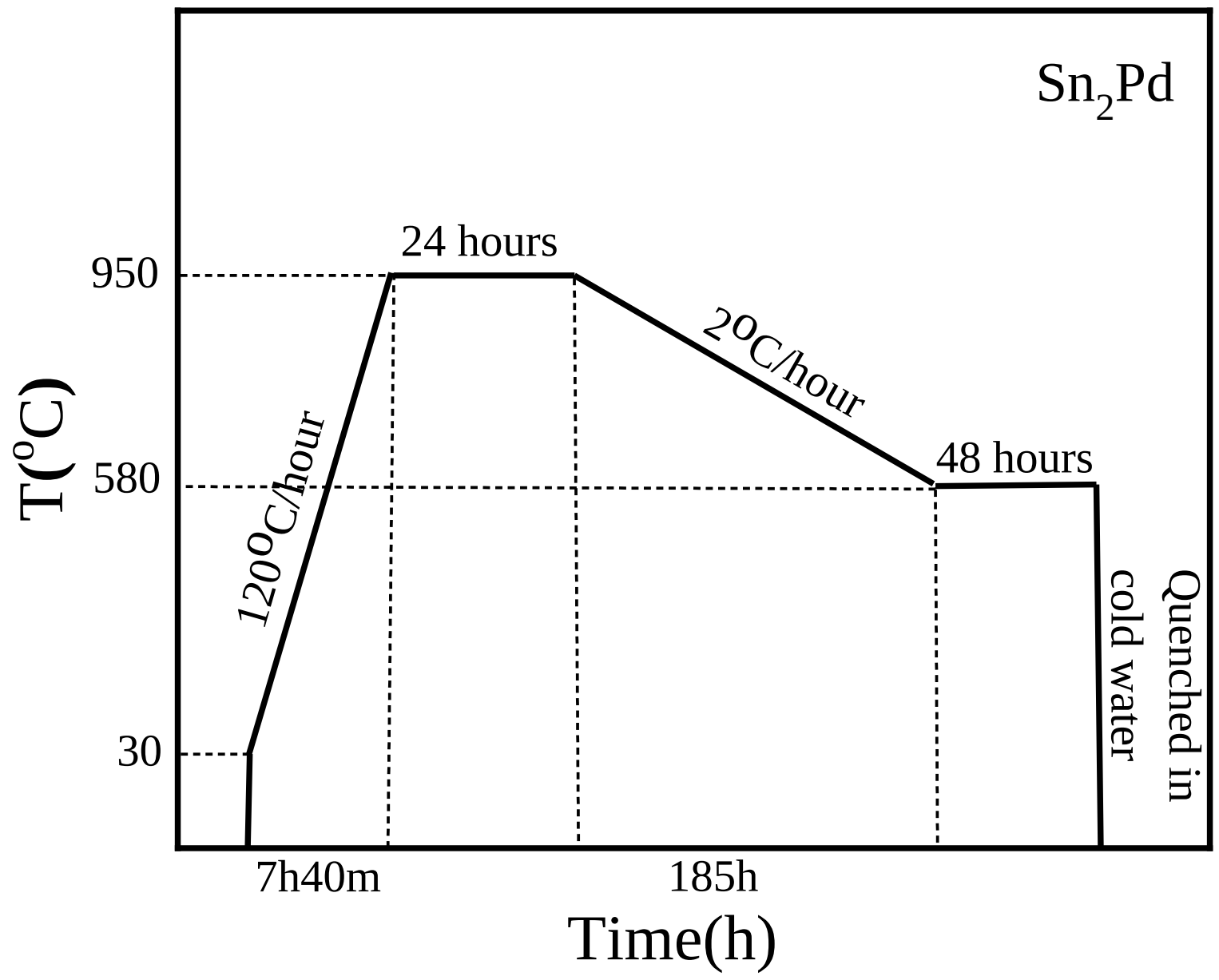

Fig. 1(b)

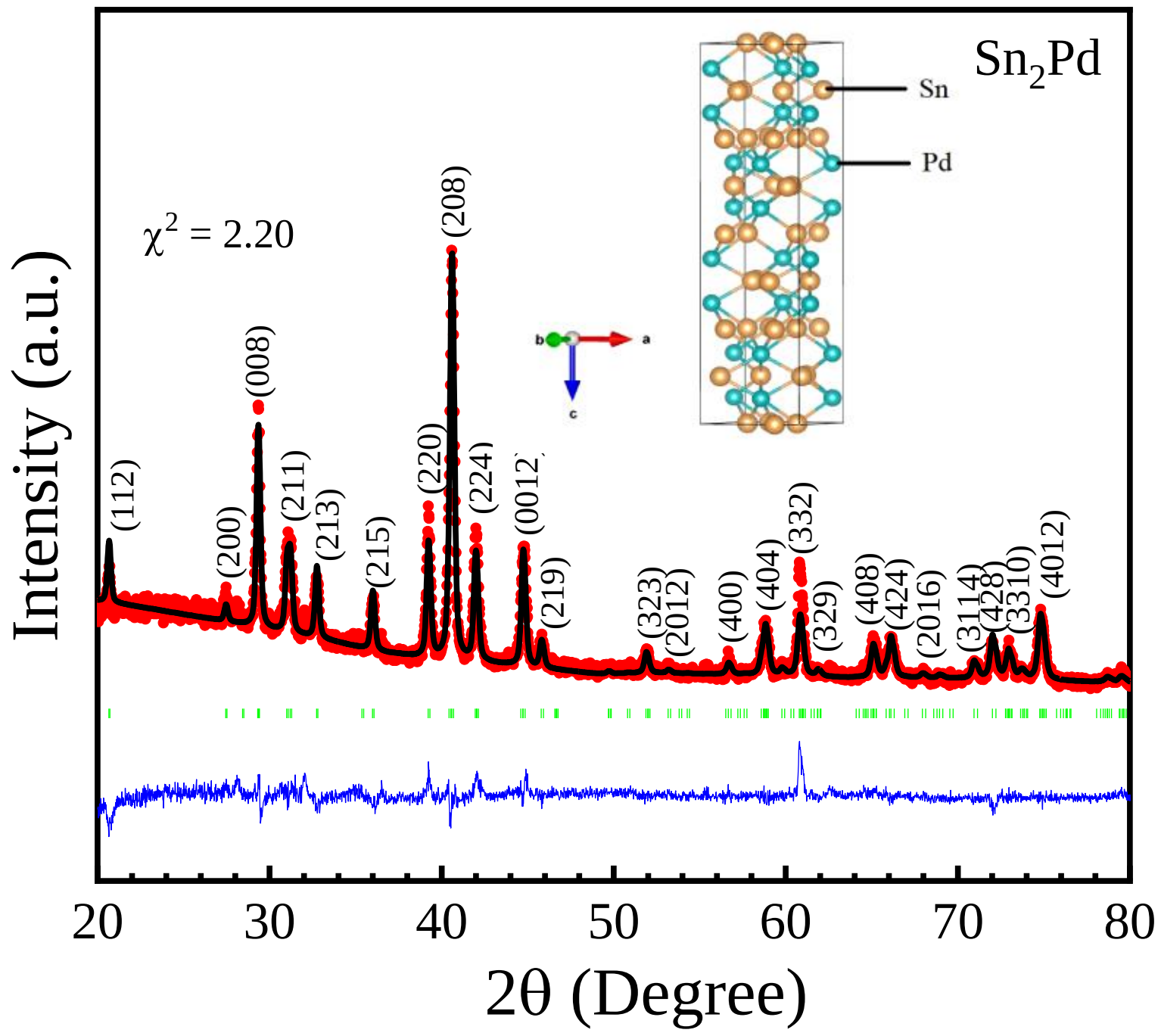

Fig. 2(a)

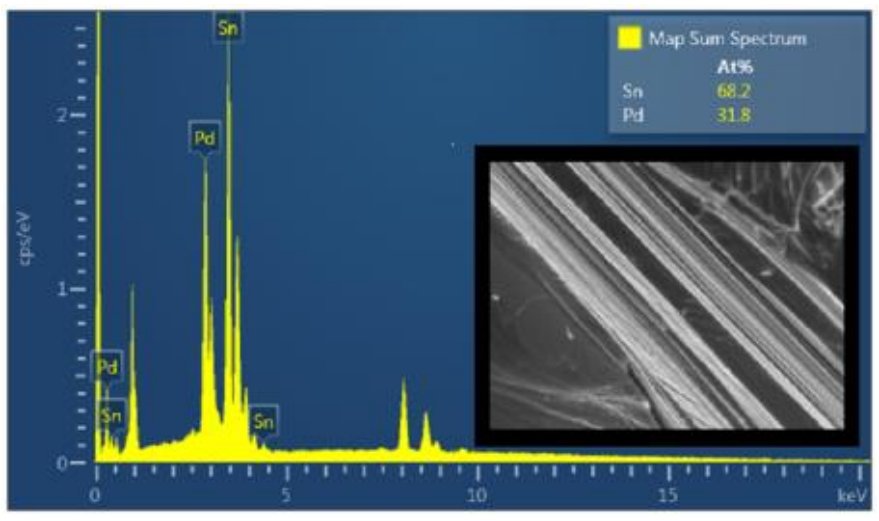

Fig. 2(b)

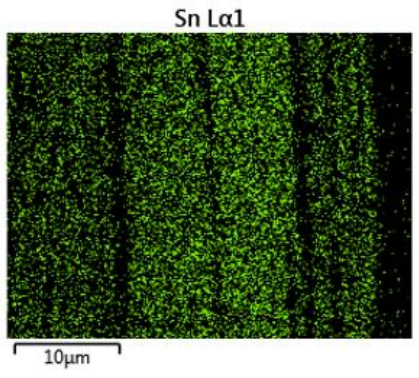

Fig. 2(c)

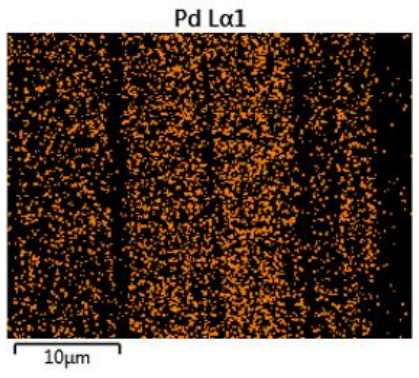

Fig. 3(a)

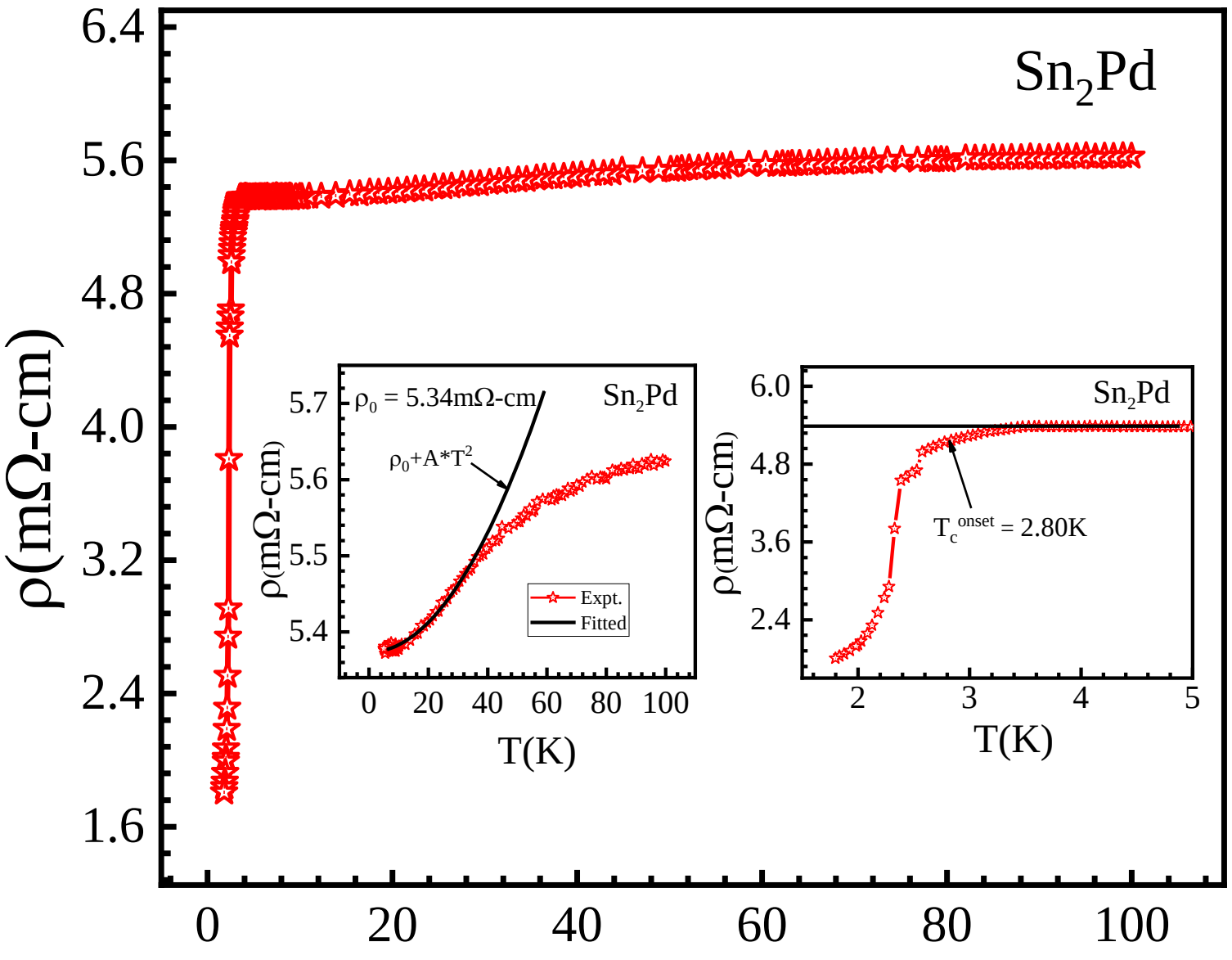

Fig. 3(b)

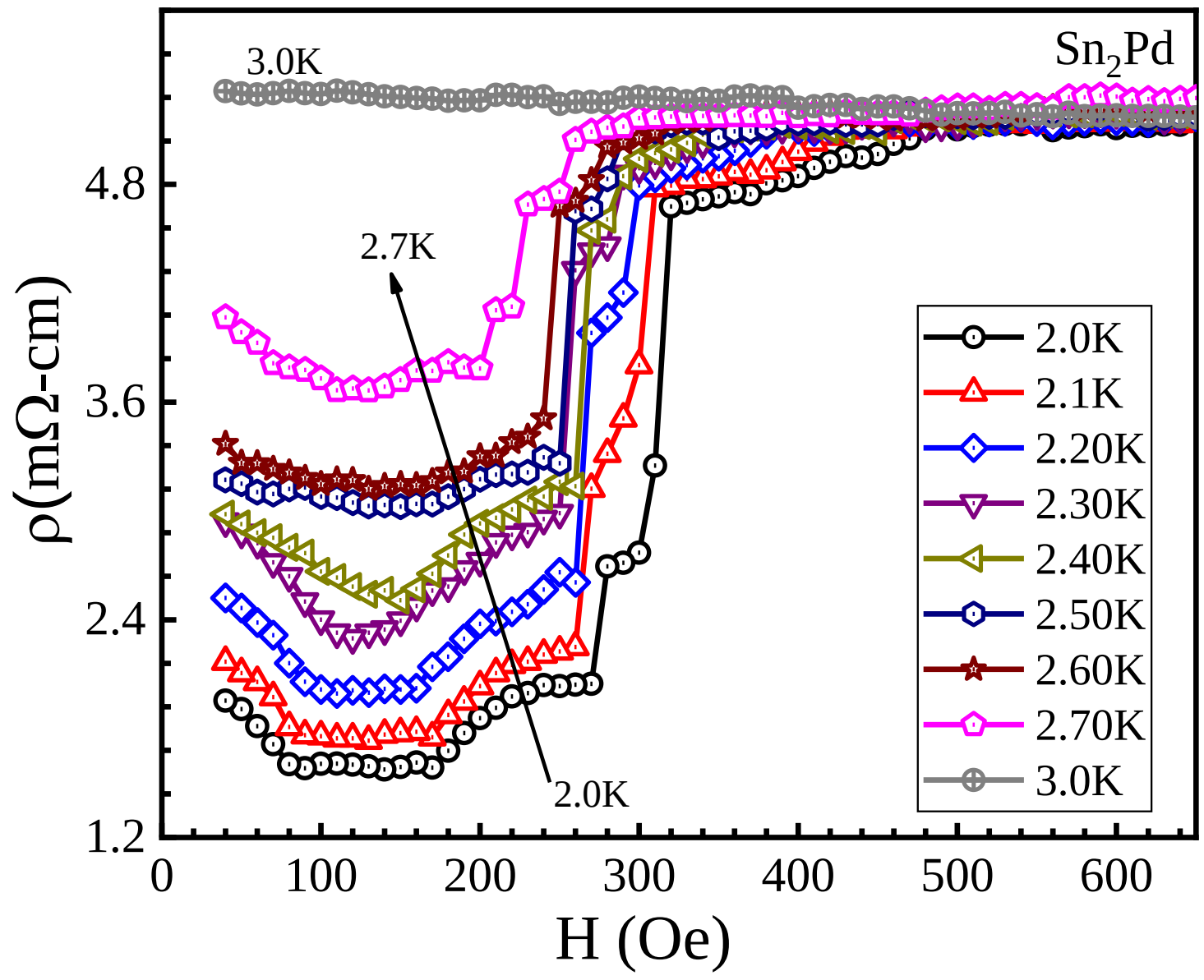

Fig. 4

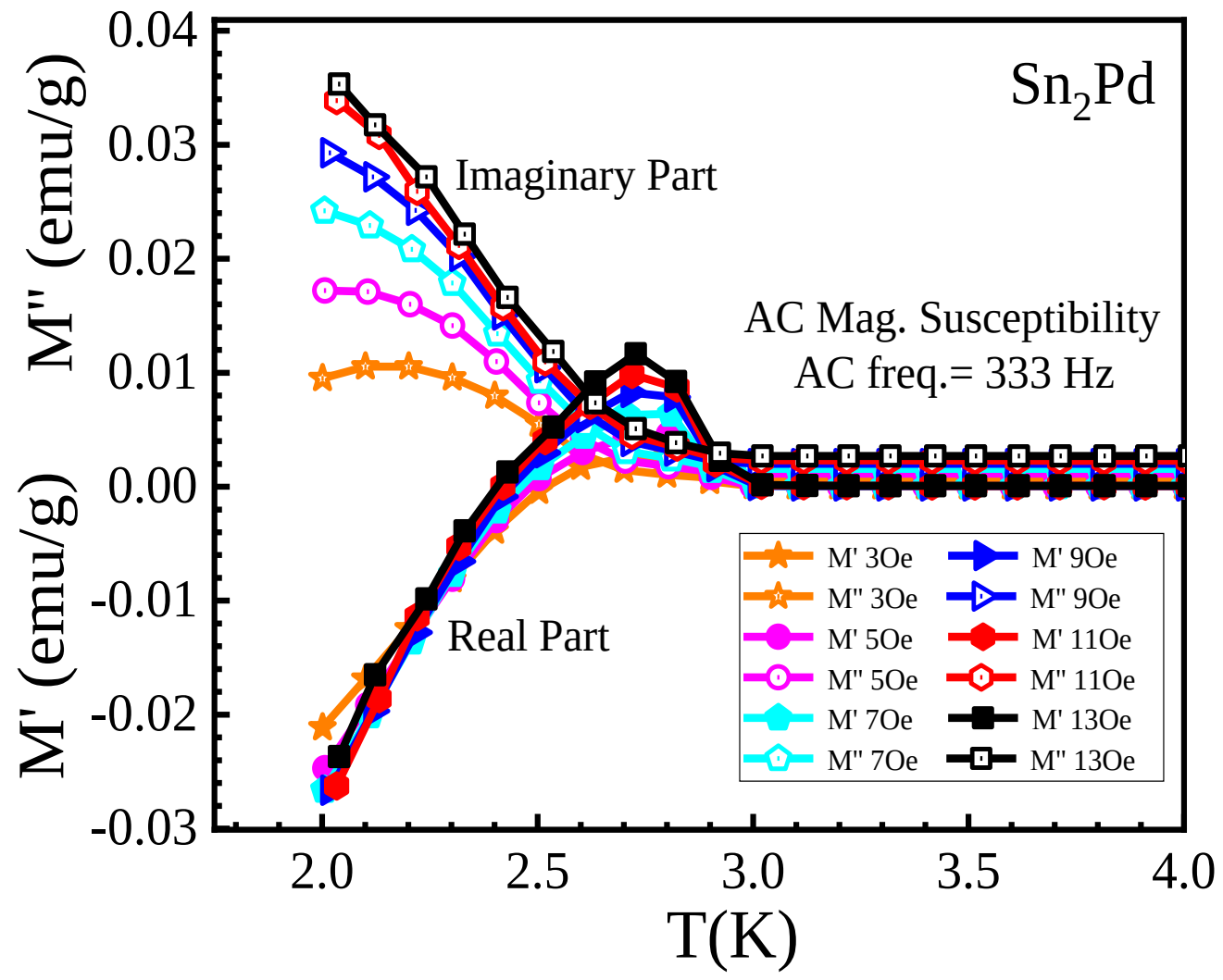

Fig. 5

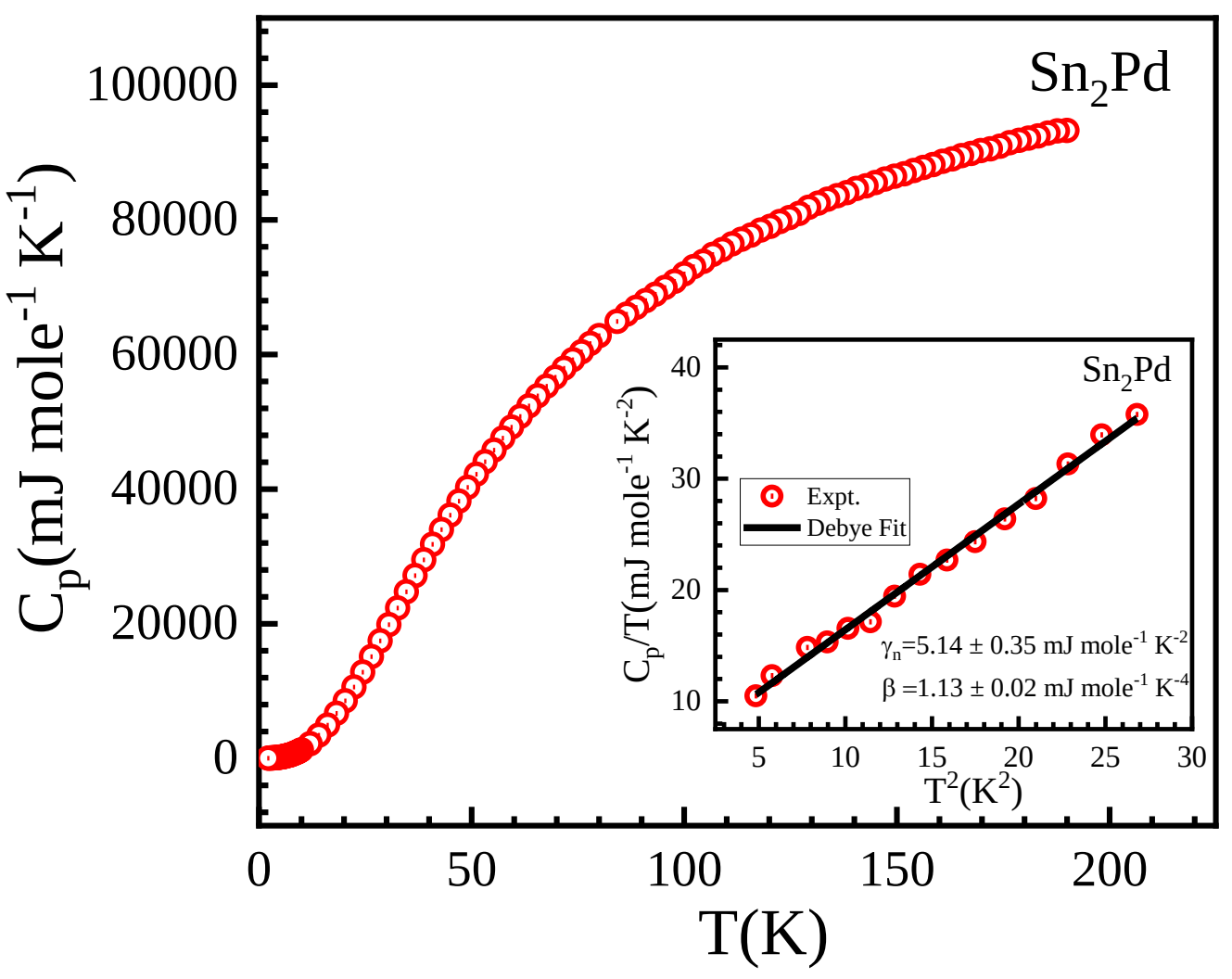

Fig. 6(a)

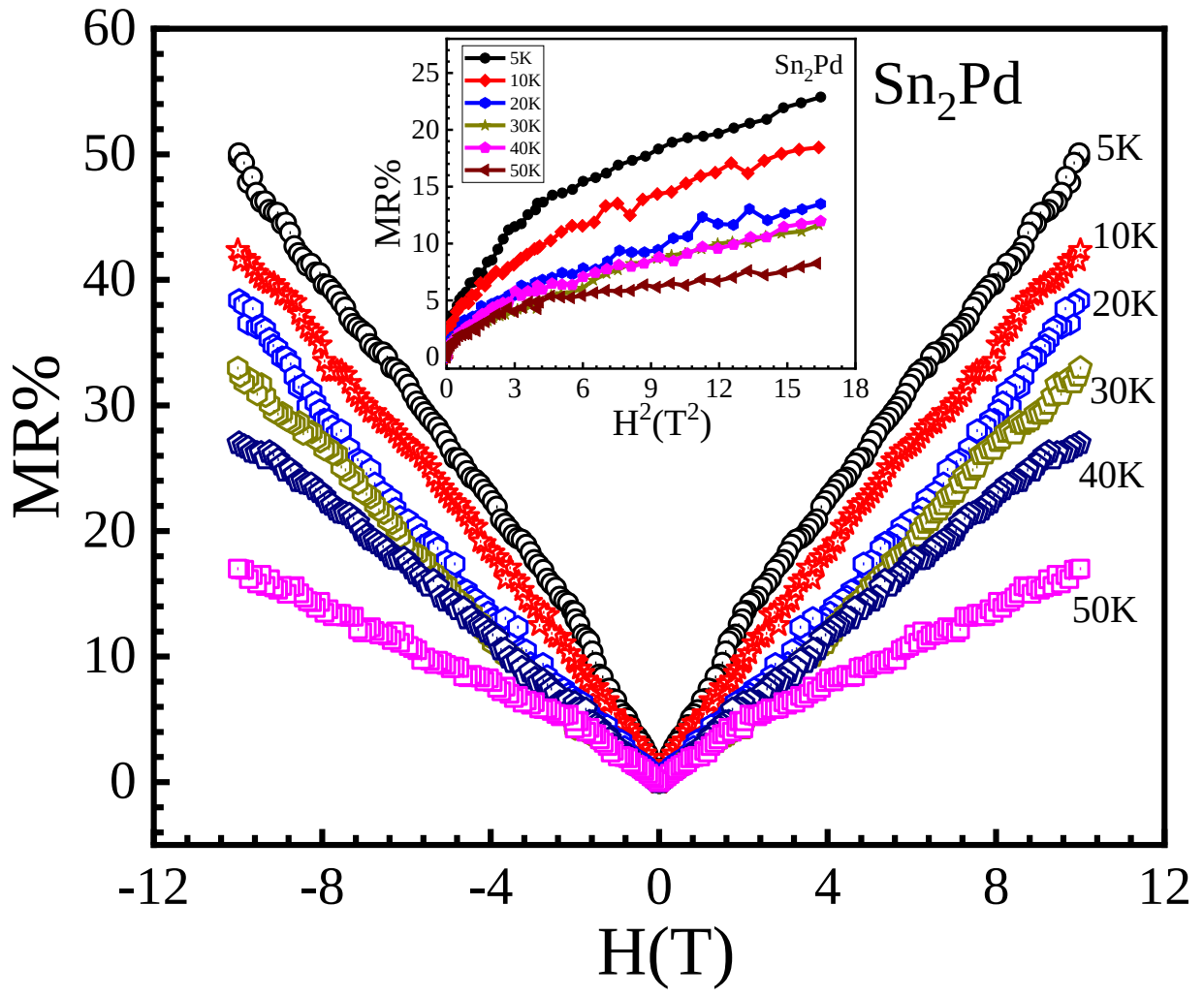

Fig. 6(b)

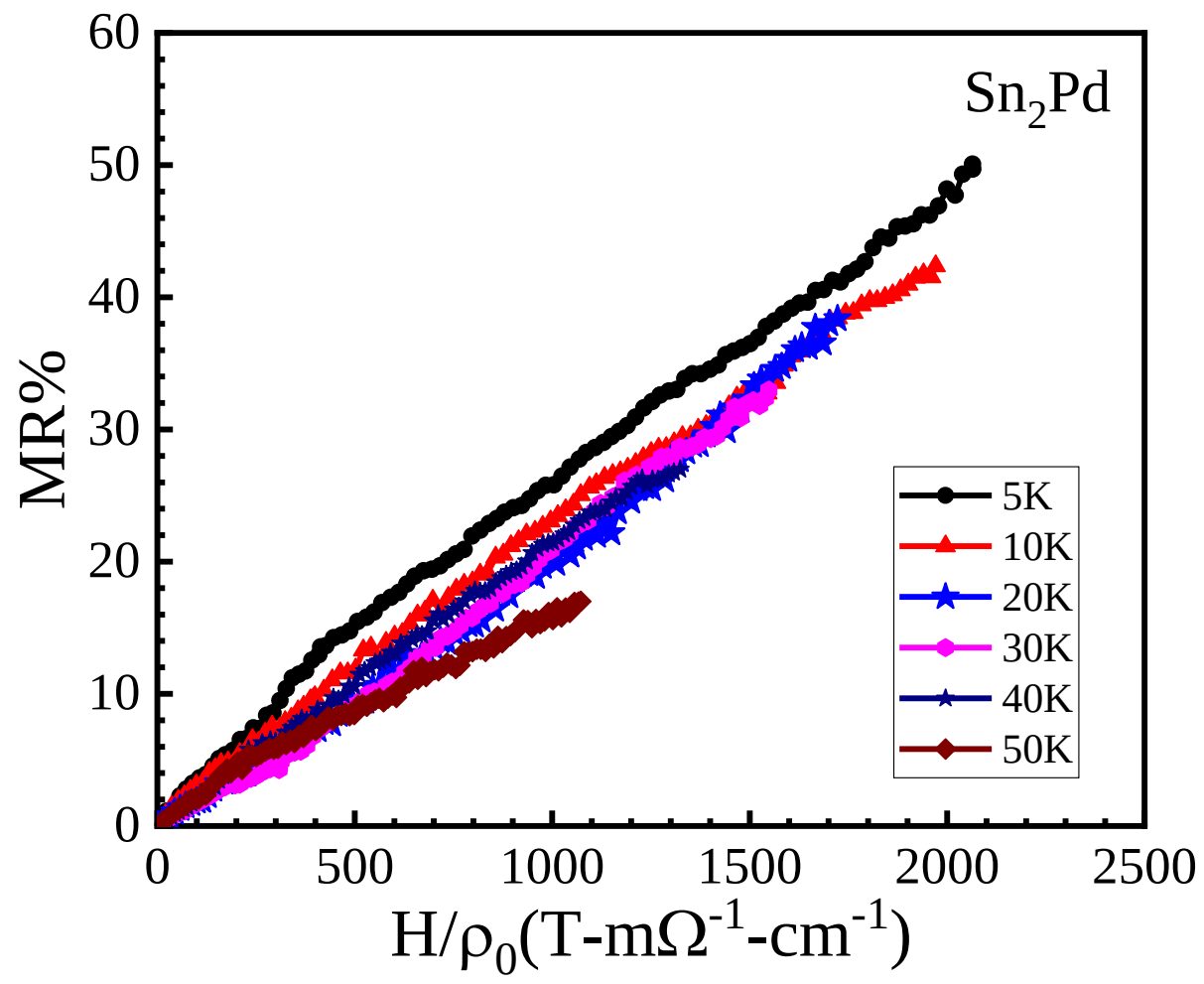

Fig. 6(c)

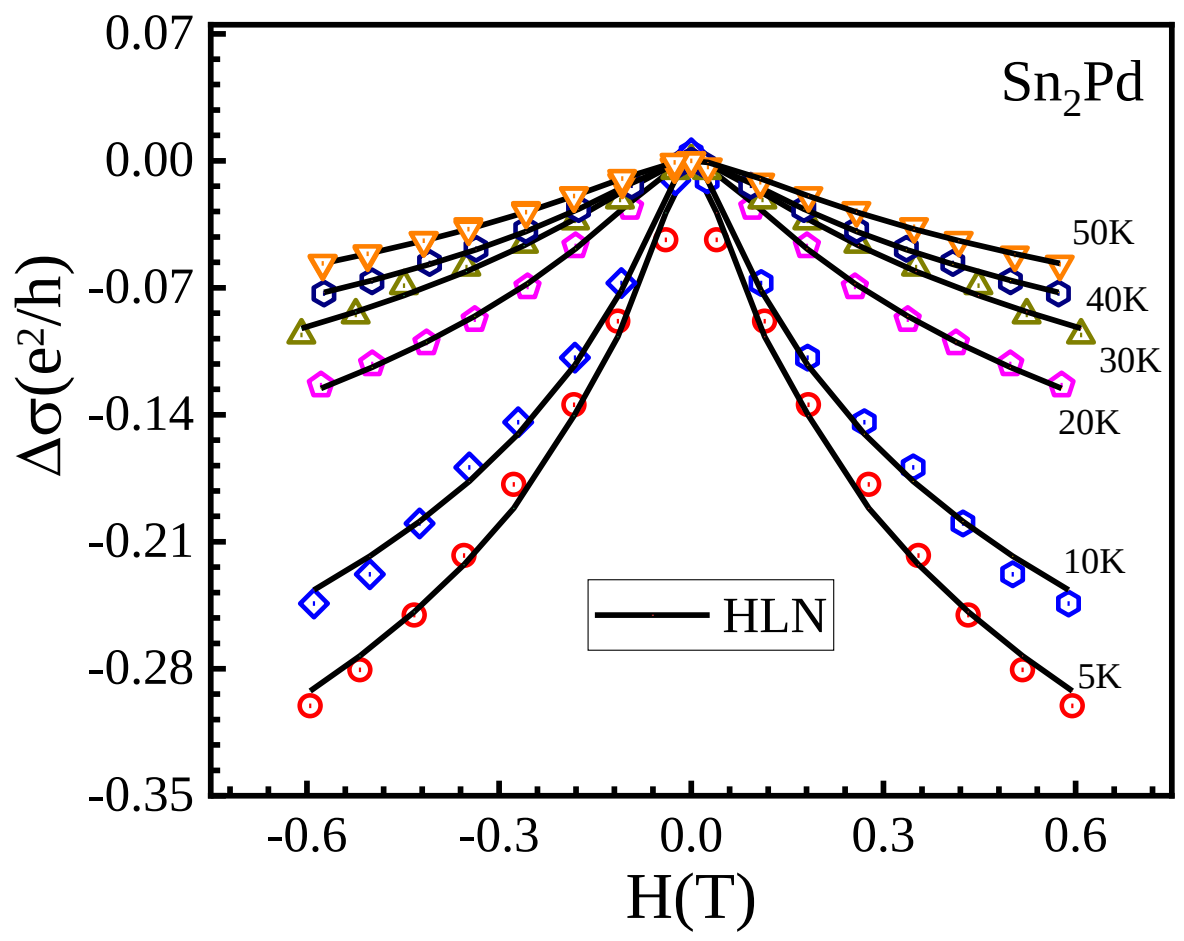

Fig. 6(d)

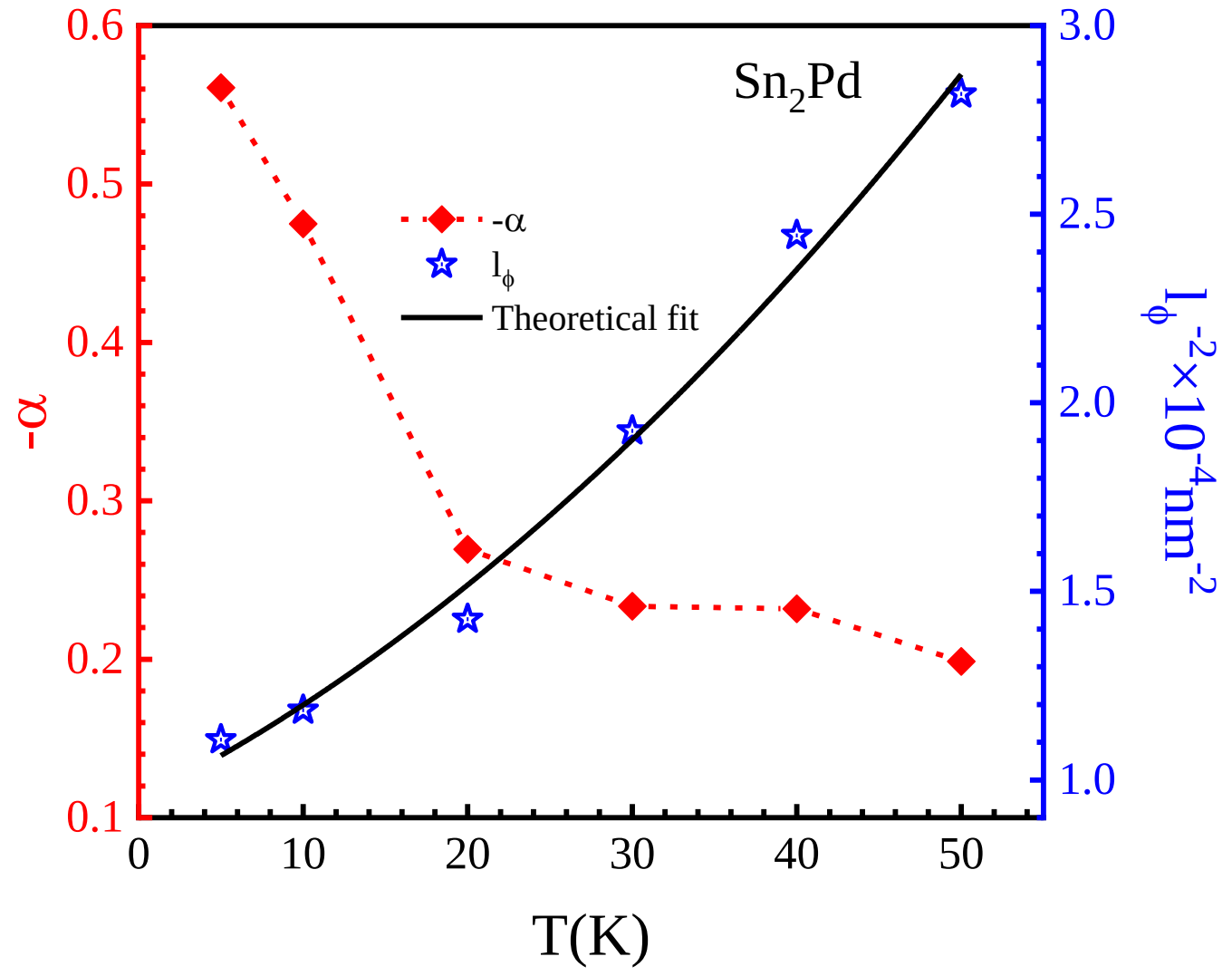